\documentclass[aps,prl,superscriptaddress,twocolumn,10pt,amsfonts,amssymb,amsmath]{revtex4}
\usepackage[pdftex]{graphicx}
\usepackage[toc]{appendix}
\usepackage[font={small,singlespacing},labelfont=bf,margin=0pt,justification=centerlast]{caption}
\usepackage[colorlinks=true,citecolor=black,linkcolor=black,urlcolor=blue]{hyperref}
\usepackage{subcaption,color,wrapfig,yllmath} 
\providecommand{\myheading}[1]{\textbf{#1}}

\newcommand{\mysubsection}[1]{ \textbf{#1:} }

\begin{document}
\title{
Superconductor-Insulator Transition and Fermi-Bose Crossovers
}

\author{Yen Lee Loh}
\affiliation{Department of Physics and Astrophysics, University of North Dakota, Grand Forks, ND 58202}

\author{Mohit Randeria}
\author{Nandini Trivedi}
\affiliation{Department of Physics, The Ohio State University, Columbus, OH  43210}

\author{Chia-Chen Chang}
\author{Richard Scalettar}
\affiliation{Department of Physics, University of California, Davis, CA 95616}

\date{\today}

\begin{abstract}
The direct transition from an insulator to a superconductor (SC) in Fermi systems is a problem of long-standing interest,
which necessarily goes beyond the standard BCS paradigm of superconductivity as a Fermi surface instability.
We introduce here a simple, translationally-invariant lattice fermion model that undergoes a SC-insulator transition (SIT) and elucidate its 
properties using analytical methods and quantum Monte Carlo simulations.
We show that there is a fermionic band insulator to bosonic insulator crossover in the insulating phase and 
a BCS-to-BEC crossover in the SC. 
The SIT is always found to be from a bosonic insulator to a BEC-like SC, 
with an energy gap for fermions that remains finite across the SIT. 
The energy scales that go critical at the SIT are the gap to pair excitations in the insulator and the
superfluid stiffness in the SC.
In addition to giving insights into important questions about the SIT
in solid state systems, our model should be experimentally realizable
using ultracold fermions in optical lattices.
\end{abstract}
\maketitle

Understanding superconductor-insulator transitions has long been an important
challenge in condensed matter physics. The Bose Hubbard model and the Josephson junction
array have led to key insights into the SIT in bosonic systems \cite{fisher1989,greiner2002,sachdev2014,auerbach2014}.
In contrast, there has been less progress in understanding the SIT in Fermi systems,
despite the existence of many electronic systems that exhibit a direct transition from an insulator to a SC.
These include, e.g., the disorder-driven SIT in thin films
\cite{haviland1989,hebard1990,yazdani1995},
 superconductivity in doped band insulators like SrTiO$_3$~\cite{behnia2013} and at
oxide interfaces~\cite{triscone2008}, and the SIT induced in Mott insulators by doping 
(high-$T_c$ cuprates)~\cite{lee2006,bollinger2011} and pressure (organics)~\cite{kanoda2011}.

Here we introduce and analyze a two-dimensional (2D) translationally-invariant lattice fermion model with local interactions 
that exhibits a direct quantum phase transition from an insulator to a SC. 
Our goal is to gain insights into a number of key issues in 
the field of SIT through a simple (disorder-free) model that can be analyzed in great detail, rather than to 
describe a specific experimental condensed matter system. 
We also note that our model can be realized experimentally using ultracold Fermi atoms in
optical lattices.

\begin{figure}
		\begin{subfigure}{.6\columnwidth}
			\includegraphics[width=\textwidth]{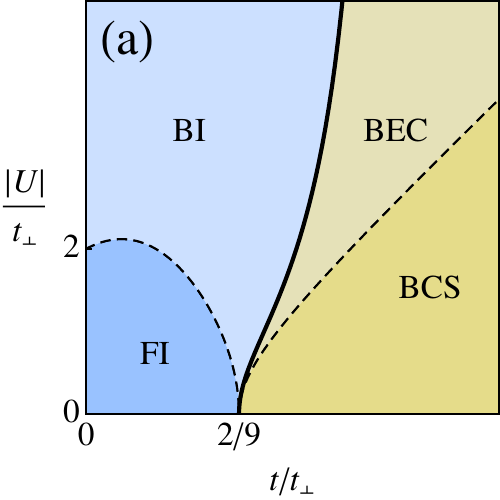}
			\label{SchematicPhaseDiagram-U-t}
		\end{subfigure}
		\hspace{.1\columnwidth}
		\begin{subfigure}{.6\columnwidth}
			\includegraphics[width=\textwidth]{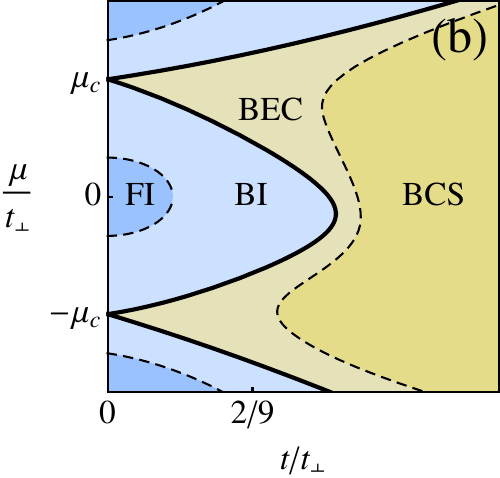}
			\label{SchematicPhaseDiagram-mu-t}
		\end{subfigure}
	\caption{
		 \label{PhaseDiagram}
		 {Schematic phase diagrams at $T=0$. (a) The $(t,|U|)$-phase diagram at density $n=1$
		 is based on mean field theory at small $|U|/t_\perp$, on QMC for intermediate $|U|/t_\perp$, 
		 on atomic limit calculations near $t/t_\perp=0$, and strong-coupling arguments at large $|U|/t_\perp$;
		 see text for details. The $|U|\!=\!0$ phase for $t/t_\perp > 2/9$ is a metal. (b) In the $(t,\mu)$-phase diagram at fixed $|U|$,
      the top, middle and bottom insulating lobes correspond to densities of $n=2,1$ and $0$ (vacuum)
      respectively. The Fermi insulator (FI)-Bose insulator (BI) crossover in (a) and (b) are defined by the nature of the excitations,
      single fermion (charge $e$) or pair (charge $2e$), with the lowest gap. The BCS-BEC crossover in the SC is determined by 
      location of the minimum gap in $\kkk$-space; see text.} 
      }
	\end{figure}

In the standard Bardeen-Cooper-Schrieffer (BCS) paradigm, superconductivity is a Fermi surface instability of a normal metal. The
challenge here is to understand how superconductivity arises out of an insulator that has no Fermi surface.
The key insight from our work is that there are Fermi-Bose crossovers in both the insulating and in the 
superconducting phases, and the SIT is always between a bosonic insulator 
and a SC in a Bose-Einstein condensation (BEC) regime; see Fig.~\ref{PhaseDiagram}.
In the weak coupling limit, though, the bosonic regimes on either side of the SIT may be narrow. 

Our main results are:
\\
1) The pairing susceptibility in the insulator diverges 
and the gap to pair excitations $\omega_\text{pair}$ in the insulator vanishes upon approaching the SIT.
\\
2) The single-particle energy gap $E_g$ for fermions remains finite in both phases across the SIT.
\\
3) The SC state is characterized by a pairing amplitude $\Delta$ and a superfluid stiffness $D_s$, 
both of which vanish approaching the SIT.  
\\
4) The insulating state near the SIT is bosonic in the sense that the gap to pair excitations is much smaller
than the fermionic gap. 
\\
5) The SC in the vicinity of the SIT is in a BEC-like regime with several unusual properties. 
Its fermionic energy gap $E_g=[(E_g^0)^2 + \Delta^2]^{1/2}$
depends upon both pairing $\Delta$ and the insulating gap $E_g^0$, and its superfluid stiffness $D_s$ is much smaller than the energy gap.
\\
6) The BCS to BEC crossover can be precisely identified by a change in the topology
of the minimum gap locus from a ${\bf k}$-space contour (BCS)
 to a point in ${\bf k}$-space (BEC).
This leads to a gap edge singularity in the fermion density of states with 
an inverse square-root divergence in the BCS regime
but a jump discontinuity (in 2D) in the BEC regime.
  
We work with a half-filled attractive Hubbard model on a triangular lattice bilayer;
the reasons for this particular choice of lattice are explained in detail below.
Our results are based on a variety of analytical 
approaches, including a strong coupling analysis about the atomic limit, 
a weak-coupling analysis of the pairing instability in an insulator and mean field theory (MFT). 
We also present numerical results from determinant quantum Monte Carlo (DQMC) simulations
that are free of the fermion sign-problem. 

Before describing our work in detail, we comment on its relationship 
with the classic paper by Nozieres and Pistolesi
on ``pairing across a semiconducting gap''~\cite{nozieres1999}.
They used MFT and estimates of phase fluctuations to analyze 
superconductivity in a system with a band gap that separates
two bands, each with a constant density of states (DOS).
Building on their ideas, our work goes beyond their analysis in terms of 
what we calculate and the methodology used, and this 
leads to new insights into the problem. 
Our explicit lattice Hamiltonian permits us to use DQMC and is
of a form that can be realized in cold atom experiments.
At the end of the paper, we comment on the connection between our results and other problems -- such as 
the disorder-tuned SIT, the superfluid-Mott transition for bosons and the BCS-BEC crossover in multi-band systems.

   \begin{figure}
      \includegraphics[width=0.99\columnwidth]{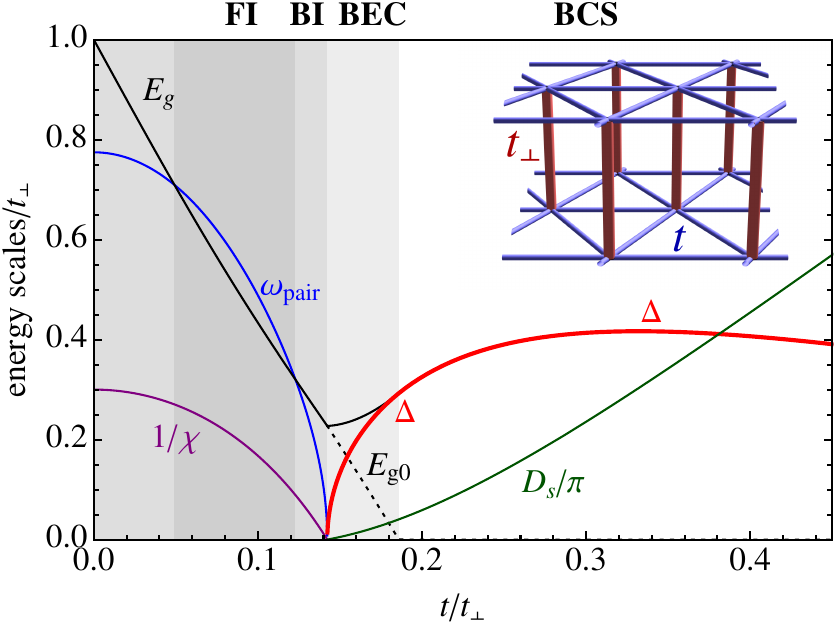}
   \caption{
      \label{EnergyScales}
      Mean-field theory (MFT) results across the $T\!=\!0$ SIT in the attractive Hubbard model on a triangular lattice bilayer (inset). 
			The SIT is tuned by $t/t_\perp$ at fixed filling $n=1$ and attraction $|U|/t_\perp=1.7$. 
			The inverse pairing susceptibility $1/\chi$ and gap to pair excitations $\omega_\text{pair}$ in the insulator
                   both vanish at the SIT.  The SC is characterized by a pairing amplitude $\Delta$ and superfluid stiffness $D_s$, which also vanish at the SIT. 
                  The single-particle energy gap $E_g$ remains finite across the SIT: $E_g=E_g^0$ in the insulator, $E_g=[(E_g^0)^2 + \Delta^2]^{1/2}$
                  in the BEC regime near the SIT and $E_g=\Delta$ is the BCS regime.
        }
   \end{figure}
   
\myheading{Model:}
We begin with the constraints on a fermion Hamiltonian that realizes a band insulator-SC transition.  
First, we need at least two sites (or orbitals) per unit cell to describe a band insulator.
Second, we must ensure that the attraction needed for SC does not
lead to other broken symmetries.
The attractive Hubbard model on a bipartite lattice has an SU(2) symmetry at half-filling,
with a degeneracy between SC and charge density waves (CDW) that leads to $T_c\!=\!0$ in 2D.
To avoid this, we choose a non-bipartite lattice.
Finally, we want to tune the SIT at a fixed commensurate filling.
Away from this filling the band insulator becomes a metal,
and we do not get an insulator to SC transition.
 
A simple model that meets these criteria is the attractive Hubbard model on two coupled
triangular lattices (inset of Fig.~\ref{EnergyScales}) with the Hamiltonian
  \begin{align}
  H &=
   -   t \sum_{\neighbor{ij}_\parallel \sigma} 
      \left( \cdag_{i\sigma} \cccc_{j\sigma} + h.c. \right)
   -   t_\perp \sum_{\neighbor{ij}_\perp \sigma} 
      \left( \cdag_{i\sigma} \cccc_{j\sigma} + h.c. \right)
   \nonumber\\&{}~~~
  - \mu \sum_{i\sigma} n_{i\sigma}
  - |U| \sum_{i} (n_{i\up} - 1/2) (n_{i\dn} - 1/2)   .
   \label{TriBiHubHamiltonian}
  \end{align}
The spin $\sigma\!= \uparrow,\downarrow$ fermion operators at site $i$ are 
$c^{\dagger}_{i\sigma}$ and $c^{\phantom{\dagger}}_{i\sigma}$, with
hopping $t$ between in-plane neighbors $\neighbor{ij}_\parallel$
and $t_\perp$ between interlayer neighbors $\neighbor{ij}_\perp$.
The chemical potential is $\mu$,  the local attraction is $|U|$, 
and $n_{i\sigma} = \cdag_{i\sigma} \cccc_{i\sigma}$.

Recently the SIT has been studied~\cite{mondaini2015}  in an attractive Hubbard model on a square lattice with near- and
next-near-neighbor hopping and a staggered (``ionic'') potential to double the unit cell. This model differs from
ours in that it has one additional parameter and exhibits CDW order in a limiting case. 
Problems related to
the SIT and the BCS-BEC crossover have also been studied in refs.~\cite{paramekanti2006,zhai2007}.
However, the specific questions we address, the observables we calculate and the methodology we use are
different from all these references.

\myheading{Non-interacting and Atomic Limits:}
We begin with exactly solvable limits in the phase diagram in Fig.~\ref{PhaseDiagram}.
First, consider the noninteracting ($U\!=\!0$) system with dispersion
  \begin{equation}
	\vare_\kkk^0 
	= -t\sum_{m=1}^{6}e^{i\left(k_x \cos\tfrac{m\pi}{3} + k_y\sin\tfrac{m\pi}{3}\right)} -t_\perp \cos k_z -\mu.
	\label{tb-dispersion}
  \end{equation}
 Here $(k_x,k_y)$ lie in the triangular lattice Brillouin zone and
$k_z$ takes on values $0$ (and $\pi$) for the bonding (and antibonding) band.
For fixed $n\!=\!1$, this implies a transition from a band insulator (gap $E_g^0 = 2t_\perp - 9t$) to a metal at $t= 2 t_\perp/9$;
see Fig.~\ref{PhaseDiagram}(a). At $U\!=\!0$
the $(t,\mu)$ phase diagram looks qualitatively similar to Fig.~\ref{PhaseDiagram}(b), except that the 
insulating ``lobes'' are triangular in shape with $\mu_c/t_\perp\!=\!\pm1$ and the SC is replaced by a metal.
(See Fig.~\ref{PhaseDiagramTightBinding} in Appendix A).

Next consider the ``atomic'' limit $t\!=\!0$, for which
the lattice breaks up into disconnected (vertical) rungs. 
We solve in Appendix B the two-site Hubbard model on a rung for arbitrary $t_\perp$ and $|U|$.
For $n\!=\!1$ there is a 
crossover from a fermionic insulator to a bosonic insulator at $|U|/t_\perp\!\simeq\!2$.
For $|U|/t_\perp\!<\!2$, the lowest energy excitation is a
single fermion (particle or hole w), and hence similar to the fermionic 
band insulator, the ground state for $U\!=\!0$.
On the other hand, for $|U|/t_\perp\!>\!2$ the lowest energy excitation is a pair of fermions.
In the large $U$ limit, the ground state is a Mott insulator of bosons, 
with one boson per rung. 
Turning on a small hopping $t$, the system is effectively described by
boson Hubbard model on a triangular lattice with one boson per ``site''.

At $t\!=\!0$ we also find that the extent of the $n\!=\!1$ insulating phase, $(-\mu_c,\mu_c)$
in Fig.~\ref{PhaseDiagram}(b), is reduced with increasing $|U|/t_\perp$. Thus the 
$n\!=\!0$ and $2$ lobes grow in size relative to $n\!=\!1$ as $|U|/t_\perp$ increases. 
We note that the description of the atomic limit phases as `insulators' is justified given that
the gap is robust to turning on a small $t\!\neq\!0$ hopping. 
The nature of the charge gap in the insulator changes across the crossover:
in the Fermi insulator it is determined by ``charge e'' excitations while in Bose insulator by
``charge 2e'' excitations, as explained above.

\myheading{Pairing Instability in the Insulator:}
We next describe a weak-coupling theory for the 
dominant instability in the band insulator as we turn on an attraction $|U|$.
The $\qqq\!=\!0$ pairing susceptibility in the ladder approximation is given by
$\chi(\omega)\!=\!\left[\chi_0^{-1}(\omega)  - |U|\right]^{-1}$, with
$\chi_0(\omega)\!=\!N^{-1}\sum_{\kkk}(1 - 2f_\kkk)/(2 \vare_\kkk - \omega - i0^+)$.
Here $N$ is the number of lattice sites, $f_\kkk$ is the Fermi function, and 
$\vare_\kkk = \vare_\kkk^0 - \mu_H$ includes the Hartree shift $\mu_H = |U|(n-1)/2$.
We analyze the problem for $n\!=\!1$, choosing $\mu$ in the insulator so that we take a trajectory 
in Fig.~\ref{PhaseDiagram}(a) that goes through the tip of the lobe.  (This choice of chemical potential is described in
Appendix D).

The divergence of $\chi\!\equiv\!\chi(\omega\!=\!0)$ at the SIT is shown in 
Fig.~\ref{EnergyScales}. We tune through the SIT by varying $t/t_\perp$, which controls the
gap in the band structure, keeping $|U|/t_\perp$ fixed.
It is energetically favorable to create pairs of particles and of holes
when the gain in pair binding energy exceeds the band gap.
This triggers the SIT, a particle-particle channel analog 
of exciton condensation in semiconductors \cite{snoke2002,eisenstein2004}.

The dynamical pair susceptibility $\chi(\omega)$ exhibits a pole at the two-particle gap $\omega_\text{pair}$. 
We see in Fig.~\ref{EnergyScales} how $\omega_\text{pair}$, 
the energy to insert a pair into the insulator, goes soft and vanishes at the SIT.
In the SC, where pairs can be inserted into the condensate at no cost, $\omega_\text{pair}\!\equiv\!0$.
We show below that the single-particle gap $E_g$ remains finite across the SIT.

\begin{figure*}
      \begin{subfigure}{.27\textwidth}
         \includegraphics[width=\textwidth]{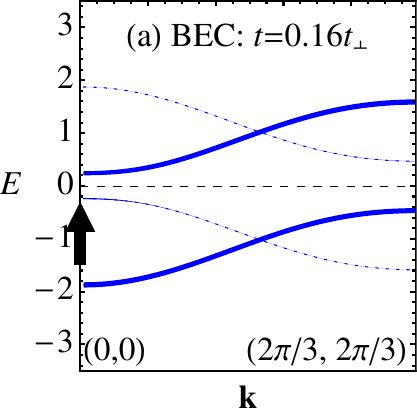}
      \end{subfigure}
      \hspace{.11\textwidth}
      \begin{subfigure}{0.27\textwidth}
         \includegraphics[width=\textwidth]{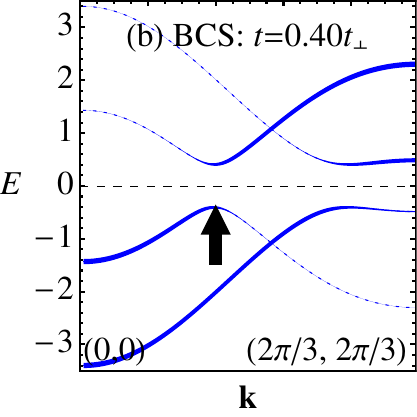}
      \end{subfigure}
      \\
      \begin{subfigure}{0.29\textwidth}
         \includegraphics[width=\textwidth]{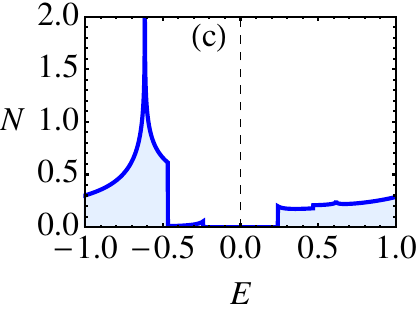}
      \end{subfigure}
      \hspace{.11\textwidth}
      \begin{subfigure}{0.29\textwidth}
         \includegraphics[width=\textwidth]{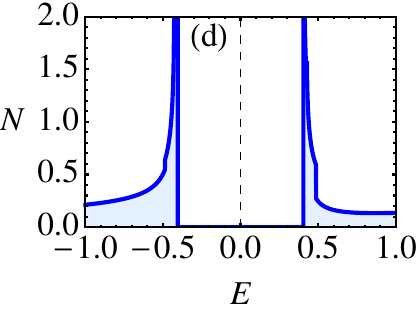}
      \end{subfigure}
   \caption{
      \label{Dispersions}
     Spectral functions (top) and density of states $N(E)$ (bottom) from $T\!=\!0$ MFT with $|U|/t_\perp = 1.7$.
      The dispersion is along the $(0,0)$ to $(2\pi/3,2\pi/3)$ in the Brillouin zone, with line thickness proportional to spectral weight.
      Left panels:  BEC regime where the lowest-energy Bogoliubov excitations occur at $\kkk=\0$, leading to a discontinuity at the gap edge in $N(E)$.
      Right panels: BCS regime where the lowest-energy excitations occur at a finite wavevector $\sim \kkk_F$, 
      leading to the usual inverse-square-root singularity at the gap edge.
      Dashed lines indicate the Fermi energy.  Arrows indicate dispersion minima.
   }
   \end{figure*}

\myheading{Mean Field Theory of SC state:}
In the small $|U|$ limit, we have a two-band superconductor, as will become apparent in the results below; see Fig.~\ref{Dispersions}(b).
We find it more convenient to analyze the problem in the site basis, rather than the band basis, given the
local attraction. Symmetry implies that the pairing amplitude is the same on the both layers, and is
defined by
$\Delta =|U|\sum_i \langle c_{i\up} c_{i\dn} \rangle/N$.
We find $\Delta$ and $\mu$ from the mean field equations 
${1}/{|U|} = N^{-1}\sum_{\kkk}  \tanh({E_\kkk}/{2T})/{2E_\kkk}$ and
$(n-1)/{2} = -N^{-1} \sum_{\kkk} ({\vare_\kkk}/{2E_\kkk}) \tanh({E_\kkk}/{2T})$.
The Hartree-shifted dispersion $\vare_\kkk = \vare_\kkk^0 - \mu_H$
determines the Bogoliubov spectrum
$E_\kkk = \sqrt{\vare_\kkk {}^2 + \Delta^2}$, from which we calculate
the energy gap $E_g = \min E_\kkk$ in the single-particle DOS $N(\omega)$.

The evolution of the $T\!=\!0$ energy gap $E_g$ across the SIT
is shown in Fig.~\ref{EnergyScales}.
In the insulator we call the gap $E_g = E_g^0$, but once 
SC sets in, we find that the SC and insulating gaps add in quadrature
$E_g=[(E_g^0)^2 + \Delta^2]^{1/2}$. For large $t/t_\perp$, the two bands merge,
the insulating gap $E_g^0$ collapses and $E_g= \Delta$, as in BCS theory. 

A single-particle gap $E_g$ that remains finite
and a pair-gap $\omega_\text{pair}$ that vanishes at the SIT implies that the insulating state close to the
SIT is a boson insulator. As discussed in the Conclusion, these results are 
very similar to those in simple models of the disorder-tuned SIT~\cite{ghosal1998,ghosal2001,bouadim2011}.

Given that the gap $E_g$ remains finite, 
what is the critical energy scale as the SIT is approached from the
SC? We show in Fig.~\ref{EnergyScales} that the superfluid stiffness $D_s$ goes soft
at the SIT. $D_s$ is obtained from the $\qqq\!\to\!0$ limit of
the transverse current-current correlation function~\cite{scalapinowhitezhang1993}.

\myheading{BEC-BCS crossover:}
We next argue that the SC state near the SIT is more akin to the BEC regime than to the BCS regime~\cite{randeria2014}. 
This is best seen from the excitation spectrum in Fig.~\ref{Dispersions}.
For large $t/t_\perp$, we are in a BCS regime, with the usual Bogoliubov dispersion 
$\pm E_\kkk$ with weights $u_\kkk^2$ and $v_\kkk^2$. The 
minimum gap $\Delta$ occurs at a finite wavevector ($k_F$ in weak coupling),
which identifies an ``underlying'' Fermi surface (FS)~\cite{sensarma2007}. 
This gap minimum located on a 1D FS contour in 2D $\kkk$-space leads to the
well known $(E-E_g)^{-1/2}$ singularity in the DOS (in addition 
to van Hove singularities in the band structure).

The BEC regime near the SIT differs from BCS in a variety of ways. 
The energy gap $E_g\!=\![(E_g^0)^2\!+\!\Delta^2]^{1/2}$
is located at $\kkk\!=\!0$. The fact that $\min E_\kkk$ occurs 
at a point (not a contour) leads to a {\it jump discontinuity} in 2D at the gap-edge (not a square-root singularity).
This qualitative difference in the DOS singularity in the BEC and BCS regimes seems not to
have been recognized earlier.

A comparison of the superfluid stiffness $D_s$ and the pairing $\Delta$ 
is also illuminating; see Fig.~\ref{EnergyScales}.
For large $t/t_\perp$ we find $D_s \gg \Delta$ as in BCS theory.  
Close to the SIT, however, we find $D_s \ll \Delta$, a BEC-like regime with well-formed pairs 
but a small phase stiffness. 

	\begin{figure}
		\begin{subfigure}{0.85\columnwidth}
			\includegraphics[width=\textwidth,trim=0 0 0 0mm,clip]{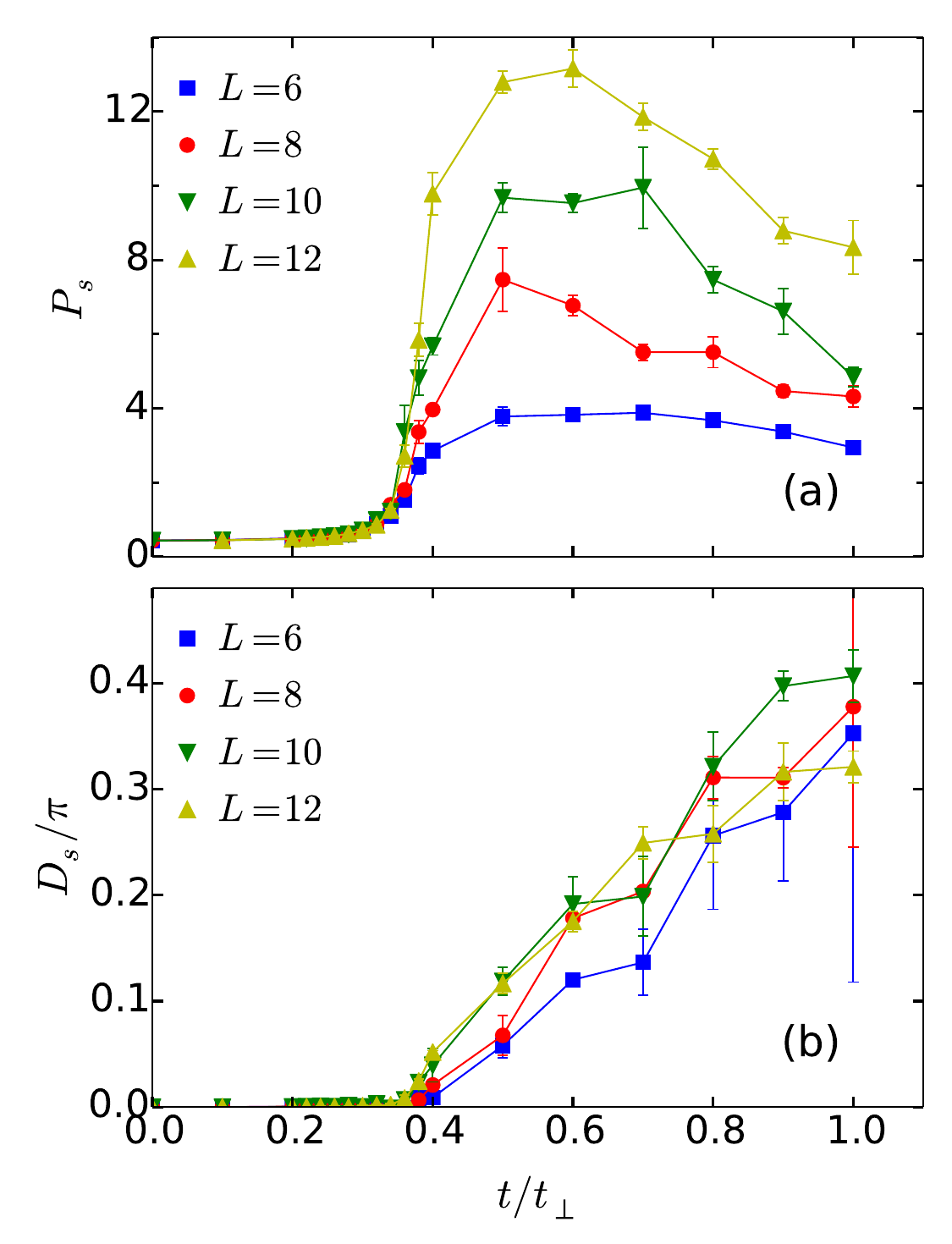}
		\end{subfigure}
	\caption{	
		\label{Dqmc}
		DQMC results across the SIT for $L\times L \times 2$ bilayer systems 
			with $|U|/t_\perp = 4$ at $T = 0.0803 t_\perp$.  
		(a) $\qqq=0$ pairing structure factor $P_s$ (see text); 
		(b) Superfluid stiffness $D_s$. 
	}
	\end{figure}
   \begin{figure}
		\begin{subfigure}{0.85\columnwidth}
			\includegraphics[width=\columnwidth]{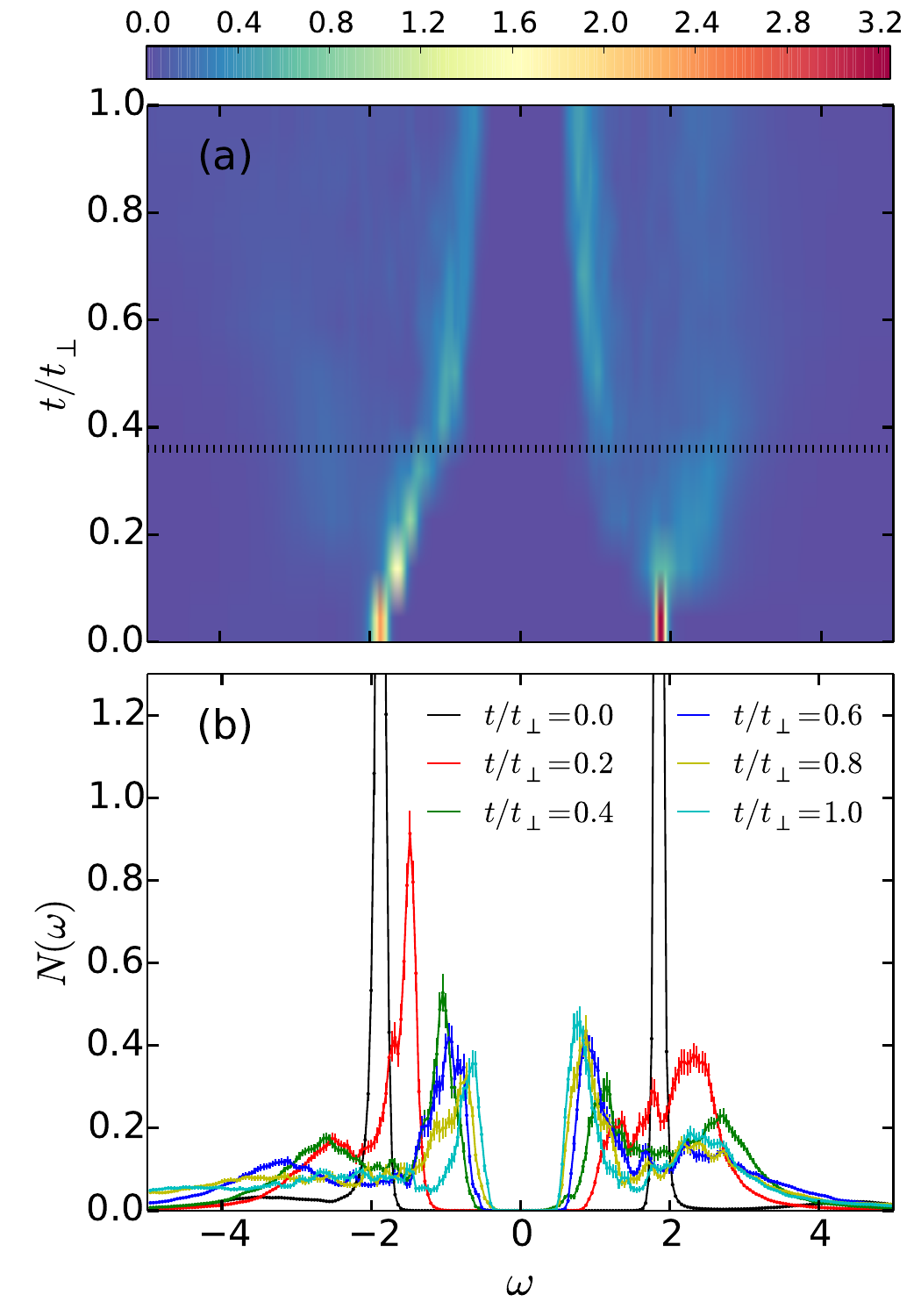}
		\end{subfigure}
   \caption{
      \label{Dos}
      DQMC density of states $N(\omega)$, calculated using the maximum entropy method,
      clearly shows the persistence of the single-particle energy gap across the SIT from the insulator to the SC.
      The dashed black line indicates the SIT.
      (a) False color plot of $N(\omega)$ as a function of $t/t_\perp$ 
      		on a $12\times 12 \times 2$ bilayer 
					with $|U|/t_\perp = 4$ and $T = 0.0803 t_\perp$;  
			(b) $N(\omega)$ for specific values of $t/t_\perp$.
   }
   \end{figure}

\mysubsection{Limitations of MFT}
{
The MFT results described above give many important insights into the SIT as a function of 
$t/t_\perp$ for fixed $|U|/t_\perp \lesssim 2$, but the approximations involved give qualitatively
incorrect results for large $|U|$. 
The mean field results of Fig.~\ref{SchematicPhaseDiagramMFT-U-t} (in Appendix D) seem to suggest
that one can induce an insulator to SC transition {\it either} by collapsing the band structure gap with
increasing $t/t_\perp$ {\it or} by increasing the attraction $|U|/t_\perp$. 
However, MFT overemphasizes order and incorrectly predicts a SC state 
at all $t/t_\perp$ for $|U|/t_\perp > 2$.

The DQMC results presented below and the atomic limit results already discussed 
give rise to the actual phase diagram shown in Fig.~\ref{PhaseDiagram}(a).
We see that one cannot go from an insulator to a SC by
increasing $U$ at a fixed (small) value of $t/t_\perp$. While pairs do form, as MFT suggests,
they do not superconduct; instead they form a bosonic insulator.
MFT fails to describe the Fermi to Bose crossover in the insulator, as well as
the SIT that occurs at $t/t_\perp$ of order unity for $|U|/t_\perp \to \infty$.
}

\mysubsection{DQMC results}
To investigate the role of quantum fluctuations beyond MFT and to obtain quantitative insights
at intermediate and large $U$, we use DQMC simulations~\cite{blankenbecler1981,hirsch1986,white1989},
to solve the triangular bilayer attractive Hubbard model, Eq.~\eqref{TriBiHubHamiltonian}.
DQMC is a statistically exact method on finite size lattices that maps the interacting electron problem
onto electrons moving in fluctuating space and (imaginary) time auxiliary fields that are sampled stochastically. 
(For QMC studies of a single triangular lattice, see Ref.~\onlinecite{nakano2006}.)

We cannot obtain $\Delta$ directly from DQMC, so we compute 
$P_s = 1/N \sum_{i,j} \langle c_{i\up}^{\phantom{\dagger}} c_{i\dn}^{\phantom{\dagger}} 
c_{j\dn}^{\dagger} c_{j\up}^{\dagger} \rangle$, the $\qqq\!=\!0$ pairing structure factor, 
which equals $\abs{\Delta}^2$ in the infinite-size limit.  
We also compute the superfluid stiffness $D_s$ using the standard Kubo 
formula~\cite{scalapinowhitezhang1993}.
We see the SIT in Fig.~\ref{Dqmc} from the onset of 
both $\abs{\Delta}^2$ and $D_s$ as a function of $t/t_\perp$ at fixed $|U|/t_\perp =4$.
We found similar results at $|U|/t_\perp=3$ (not shown).

We see that $D_s$ increases monotonically with $t/t_\perp$
but $P_s$ exhibits non-monotonic behavior, similar to the 
MFT results for $\Delta$, which we can understand as follows. 
In the BEC regime, close to the SIT, the order parameter $\Delta$ increases with $t/t_\perp$, 
however, eventually the increase in bandwidth 
leads to a smaller normal-state DOS, and the BCS $\Delta$ decreases.

Determining the universal critical exponents at the SIT would require careful
finite size scaling of the DQMC data, which is beyond the scope of this paper, where
we focus on establishing the nature of the phases. Note that 
the MFT results in Fig.~\ref{EnergyScales} exhibit mean field exponents
$\Delta \sim \delta^{1/2}$ and $D_s \sim \delta$, where $\delta$ is the deviation 
of the tuning parameter $t/t_\perp$ from its critical value.

Finally, we show in Fig.~\ref{Dos} that the persistence of a finite single-particle gap $E_g$ 
from the insulator to the SC is not an artifact of MFT, and is clearly seen in the DQMC results
for the DOS. The DOS $N(\omega)$ was obtained from analytic continuation of 
DQMC data using the maximum entropy method~\cite{sandvik1998}.  
A detailed analysis of our DQMC results will be presented elsewhere. 

\mysubsection{Comparison with other SIT problems}
We compare the results and insights obtained above with
other systems that exhibit an SIT.
Specifically, we discuss (i) the Mott insulator to superfluid transition in the
boson Hubbard model (BHM), and (ii) the SIT in a 
system of disordered fermions.

(i) The SIT in a fermionic system discussed here
shows surprising similarities with the boson superfluid-Mott transition~\cite{fisher1989}  
experimentally realized in optical lattices~\cite{greiner2002}.
The critical behavior at the SIT in our model is expected 
to be in the same universality class as the BHM.
It is, however, remarkable that the phase diagrams of the two models
in the $(\mu,t)$-plane are also similar [see Fig.~\ref{PhaseDiagram}(b)] even though,
for small $|U|$, our fermionic system {\it cannot} be mapped onto bosons.

In the large $|U|$ limit, an important difference from the standard BHM
is that our model maps on to a system of bosons with density $n_b\!=\!1/2$ per site 
(corresponding to the fermion density of $n\!=\!1$). 
Naively, one might have thought that one needs an integer boson filling to 
obtain an insulator. However our analysis shows that our large $|U|$
bosons live on rungs rather than on sites, and hence the bilayer model 
does have a bosonic insulating phase.

(ii) Next we compare the results of our disorder-free (``clean'')
fermonic SIT with the well-known problem of strongly disordered 2D SC's.
There is a large experimental literature on
superconducting films~\cite{haviland1989,hebard1990,yazdani1995,ci-qpt2012} where 
quantum phase transitions are observed as a function 
of increasing disorder or magnetic field. Broadly, the 
experiments fall into two distinct classes. Either they show
(a) a direct transition from a superconductor (SC) to an insulator, or
(b) an intermediate metallic state between the SC and insulator.
The existence of a metallic state in a 2D interacting, disordered system 
is not understood at present.
We focus here only on the direct SIT in (a).
 
One of our main motivations was 
 to find a simple (disorder-free) model that would give important insight
 into the results obtained previously on 
 SIT driven by disorder. We had previously analyzed the SIT in the
 2D square lattice attractive Hubbard model with a random on-site potential
 \footnote{In the presence of disorder, the details of the lattice are unimportant.  In particular, it does not matter whether the lattice is bipartite (e.g., square lattice) or not (e.g., triangular bilayer).}
using spatially inhomogeneous Bogoliubov-de Gennes MFT~\cite{ghosal1998,ghosal2001} and
sign-problem-free DQMC simulations~\cite{trivedi1996,scalettar1999,bouadim2011}.
These papers led to a number of striking predictions, such as
the persistence of the single-particle (fermion) gap across the SIT~\cite{ghosal1998,ghosal2001,bouadim2011}
and the existence of a pseudogap above $T_c$ in the highly disordered SC~\cite{bouadim2011},
which have been verified by scanning tunneling spectroscopy 
experiments~\cite{sacepe2010,sacepe2011,mondal2011} on disordered films of TiN and InO$_x$.
Other predictions, such as the collapse of the pairing gap in the highly
disordered insulator~\cite{bouadim2011}, have not yet been experimentally tested.

From a theoretical point of view, it is useful to ask
which of these results is special to disordered systems and 
which might be more general features of any direct SIT. 
We now discuss how the new results on the ``clean'' problem
presented in this paper shed light on these questions.

Clearly, the same response functions signal the SIT 
in both the clean and disordered problems. The
pair susceptibility diverges on approaching the transition from the insulator,
while the superfluid density vanishes on approaching the SIT from the SC side. 
The nature and evolution of the excitation spectrum is nontrivial.
Both in the clean and the disordered problems, we find that
the single-particle (fermion) gap remains finite across the transition, while
the two-particle gap collapses upon approaching the SIT from
the insulator. While this is what we would expect for a 
``bosonic'' SIT, the microscopic mechanisms underlying the 
gaps are quite different in the clean and disordered cases.

The origin of the insulating gap~\cite{ghosal1998,ghosal2001,bouadim2011} 
in the disordered attractive Hubbard model crucially involves the {\it spatial inhomogeneity} of
the pairing amplitude, and leads to an unusual insulator with localized pairs. 
The fact that the single particle gap in the disordered insulator
remains ``hard'' (i.e., zero spectral weight within the gap at 
$T=0$), despite the Griffiths' effects of rare regions, is a subtle and
surprising effect~\cite{loh2016}. In contrast, the persistence of the fermion gap 
across the SIT in the clean problem can be  
traced to the band gap in the band insulator. This important 
insight is made quantitative in the small $|U|$ limit, where we have
derived a simple analytical expression
that shows how the insulating and SC gaps add in quadrature.

We have identified: (i) a two-particle gap $\omega_{\rm pair}$ that is lower than the single-particle gap
as the defining characteristic of a ``bosonic insulator'', and (ii) the vanishing
of $\omega_{\rm pair}$ as characteristic of a bosonic SIT. 
In the disordered case,  $\omega_{\rm pair}$ was obtained~\cite{bouadim2011} 
only after considerable numerical effort  by analytically continuing imaginary-time DQMC data to real frequencies.
In addition, Griffiths' effects due to rare regions give rise to spectral weight at arbitrarily low
energies in the two-particle spectral function~\cite{bouadim2011,swanson2014}. Thus 
$\omega_{\rm pair}$ is a well-defined scale, but not a hard gap in the disordered case.
In contrast, the two-particle gap $\omega_{\rm pair}$ computed in this paper is a hard gap that can be 
analytically obtained (at least in the small $|U|$ regime) in a diagrammatic approach,
only because of the simplicity of the model introduced here.

The BCS to BEC crossover is a well studied problem~\cite{randeria2014};
(see below for comments on the multi-band case). To the extent that the superfluid stiffness is 
much smaller than the energy gap, the SC close to the SIT is in a BEC regime in
both the disordered and clean problems. A new insight obtained in the clean problem
is the change in topology of the minimum gap locus from a ${\bf k}$-space contour in the BCS regime
(the Fermi surface in the weak coupling limit) to a point in ${\bf k}$-space in the BEC regime. This immediately leads
to a change in the gap edge singularity from an inverse square-root divergence in the BCS regime
to a jump discontinuity (in 2D) in the BEC regime. These sharp diagnostics are not available
in the disordered case where ${\bf k}$ is not a good quantum number.

The Fermi to Bose crossovers in the insulating state and the microscopic nature of the bosonic insulator near the 
SIT are different in the clean and disordered cases. In the disordered problem, one
has an Anderson insulator in the non-interacting Fermi limit, but
a non-trivial and highly inhomogeneous state of localized pairs in Bose insulating regime
close to the SIT~\cite{ghosal1998,ghosal2001,bouadim2011}.
On the other hand, in the clean problem analyzed here, the Fermi insulating regime
is smoothly connected to a non-interacting band insulator, while the Bose insulator is
smoothly connected to a Mott insulator.

In summary, both the clean and the disordered problems have a bosonic SIT
with bosonic phases (Bose-insulator and BEC) on either side of the transition.
Nevertheless, as described above,
the microscopic manner in which a system of fermions leads to
bosonic physics is completely different in the clean and disordered systems. 
The disorder-free model introduced
and analyzed here has the virtue that we can obtain analytical insight, supplemented 
by DQMC solutions in the intermediate coupling regime. 
The disorder-driven SIT is a more complex problem with
strong disorder and interactions, 
which cannot be analyzed in as transparent a fashion as the 
translationally invariant problem of this paper.

\mysubsection{Experimental Implications and Conclusions}
We have addressed in this paper the conceptual question of 
how superconductivity develops in an insulator using a simple fermonic model.
We conclude with the experimental implications of the
the insights gained from our analysis. 

All of our results can be {\it quantitatively} tested, in principle, 
in optical lattice experiments with cold atoms~\cite{esslinger2010}
A triangular lattice bilayer can be made using existing optical lattice
techniques. There are established ways of probing both the insulating and 
the superfluid phases, and measuring their excitation spectra. 
The most important experimental challenge here would be 
to cool fermions in an optical lattice below their superfluid transition.
This issue is being intensively pursued at the present time. For recent 
experimental progress on the emulating the repulsive Hubbard model in an optical lattice, 
and comparisons with QMC, see ref.~\onlinecite{hulet2015}.

Perhaps even more interesting than the direct experimental test of the specific model that
we analyze, are the {\it qualitative}, model-independent insights 
on the Fermi-Bose crossovers in the insulator, the bosonic nature of the transition 
and the results (1) through (6) highlighted in the Introduction. There are several 
classes of materials for which these could be relevant. 

Here we briefly discuss
the case of superconductivity in bulk FeSe and FeSeTe, where there is considerable 
evidence~\cite{lubashevsky2012,kasahara2014,okazaki2014}, 
for a single-particle energy gap comparable to the Fermi energy, a hallmark of 
the crossover regime in between the BCS and BEC limits.

Previous investigations of the BCS-BEC crossover problem~\cite{randeria2014}
have focused almost exclusively on single-band systems in the context of ultracold Fermi gases. 
On the other hand, the bulk FeSe materials are compensated semi-metals, where 
it is essential to take into account multiple bands. The study of the BCS-BEC crossover in 
multi-band systems is in its infancy. While not a microscopic model
for FeSe, the model presented here is a compensated two-band semimetal that exhibits 
the BCS-BEC crossover. Hence, the qualitative insights derived from our work are clearly relevant. 
An important prediction of our analysis is the 
change in the topology of the minimum gap locus as one goes from the BCS to the BEC regime
and the corresponding change in the gap-edge singularity in the DOS. 
These could be very useful diagnostics for identifying the BCS and BEC regimes in 
the SC state of bulk FeSe, a problem worthy of detailed investigation in the future.
\bigskip

\myheading{Acknowledgments:}
We gratefully acknowledge support from NSF DMR-1410364 (MR), DOE DE-FG02-07ER46423 (NT), 
and from the UC Office of the President (CC, RTS). 
M.R. would like to acknowledge very useful
feedback from Antoine Georges and Doug Scalapino on a preliminary version of this paper.

\bigskip\bigskip

\appendix

\setcounter{section}{1}  
\renewcommand{\theequation}{\Alph{section}\arabic{equation}}

\section{Appendix A: Non-interacting limit}
First consider the non-interacting tight-binding model on a triangular bilayer
(Eq.~\eqref{TriBiHubHamiltonian} with $\abs{U}=0$), whose dispersion relation is given by Eq.~\eqref{tb-dispersion}.
The density of states (DOS) is
	\begin{align}
	g(E)
	&= \frac{1}{2t} \left[
		g_\text{tri} \left( \frac{E - t_\perp}{t} \right) +
		g_\text{tri} \left( \frac{E + t_\perp}{t} \right)
	\right]	
	\label{TriBiDOS}
	\end{align}	
where 
	$g_\text{tri} (\vare)
	=	-\frac{1}{\pi} \Im G_\text{tri} (\vare+i0^{+})
	$
is the density of states of a \emph{single} triangular lattice, and $G_\text{tri}$ is the triangular lattice Green function \cite{horiguchi1972},
	\begin{align}
	G_\text{tri}(\vare)
	&=
		\frac{1}{\pi  \left(\frac{\vare }{2}-1\right)^{3/4} \left( \frac{\vare }{2}+3 \right)^{1/4} }   
			\times\nonumber\\&~~~~{}
		K\left(\frac{
				\frac{\vare ^2}{4}-3}{2 \left(1-\frac{\vare }{2}\right) 
				\left(  \frac{\vare }{2} - 1 \right)^{1/2}
		  	\left(  \frac{\vare }{2}+3   \right)^{1/2}   
			}+\frac{1}{2}\right)
	,
		\label{TriLattGF}
	\end{align}
where $K$ is the complete elliptic integral of the first kind as implemented by the Mathematica function \texttt{EllipticK}.

\begin{figure*}[!htb]
      \begin{subfigure}{0.26\textwidth}
         \includegraphics[width=\textwidth]{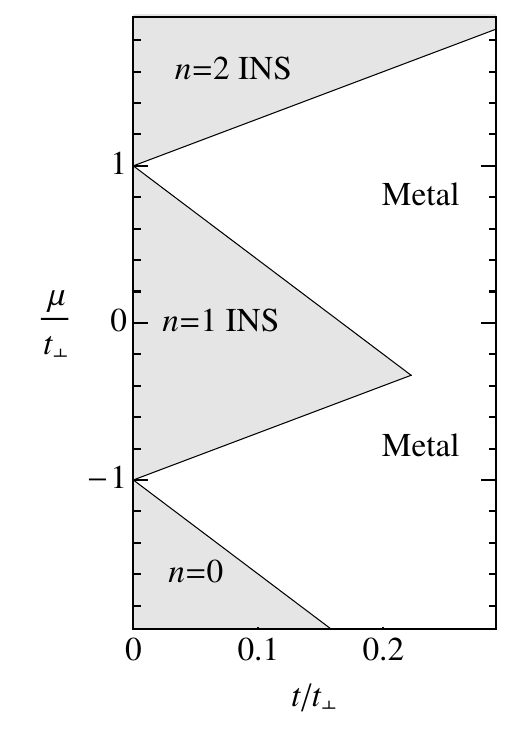}
         \caption{
	         \label{PhaseDiagramTightBinding}
         }
      \end{subfigure}
      \begin{subfigure}{.26\textwidth}
         \includegraphics[width=\textwidth]{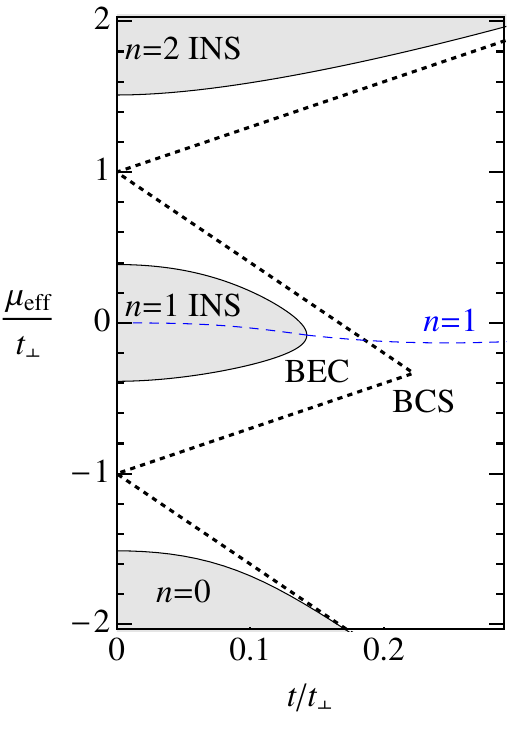}
         \caption{
	         \label{PhaseDiagramMuEff}
         }
      \end{subfigure}
      \begin{subfigure}{.26\textwidth}
         \includegraphics[width=\textwidth]{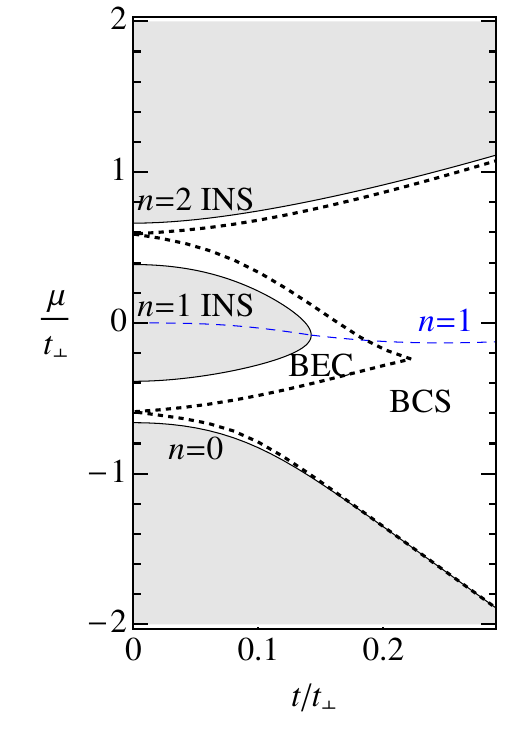}
         \caption{
	         \label{PhaseDiagramMu}
         }
      \end{subfigure}
   \caption{
      \label{adsf}
			(a) Phase diagram of {\it non-interacting} triangular bilayer tight-binding model.
					As $t$ increases, the valence band and conduction band broaden
						and eventually overlap, causing a $n=1$ band insulator to
						metal phase transition.
			(b)	Phase diagram of attractive Hubbard triangular bilayer
					 as function of $\mu_\text{eff}$ at $|U|/t_\perp=1.7$,
					where $\mu_\text{eff}$ includes the Hartree shift.
					The triangular regions shrink to lobes
						at the locus of the pairing instability.
					A portion of the insulating phase in (a) phase is converted into a BEC superfluid,
						whereas the metallic phase turns into a BCS superfluid.
			(c)	Phase diagram of attractive Hubbard triangular bilayer
					 as function of actual chemical potential $\mu$ at $|U|/t_\perp=1.7$.
					The $n=2$ insulating lobe is displaced downward by $\abs{U}/2$,
					whereas the $n=0$ lobe moves up by $\abs{U}/2$.
					In the superfluid, lines are distorted due to Hartree shifts
						by non-integer $n$.
   }
   \end{figure*}

For small $t$, this model possesses a valence band with energies 
$-t_\perp-6t \leq E \leq -t_\perp+3t$ 
and a conduction band with energies
$+t_\perp-6t \leq E \leq +t_\perp+3t$.
When $t > 2t_\perp/9$ the bands overlap to form a single band.
The model behaves as a band insulator or metal depending on where the Fermi energy $\mu$ lies within the bandstructure.  
This leads to the phase diagram in Fig.~\ref{PhaseDiagramTightBinding}.

\section{Appendix B: Atomic Limit}

\begin{figure}[!h]
	\includegraphics[width=0.8\columnwidth]{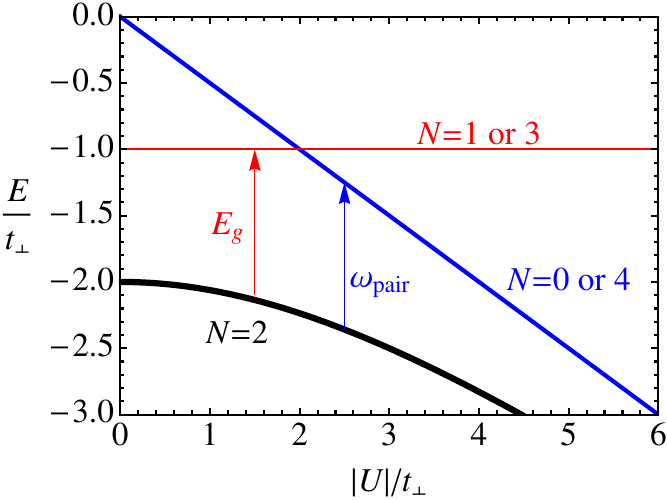}
	\caption{
	\label{AtomicLimitEnergyLevels}	
		Energy levels for a two-site Hubbard model as a function of $U$, for $\mu=0$.
	}
	\end{figure}

Now consider the triangular bilayer Hubbard attractive model in the ``atomic limit'' ($t=0$, $\abs{U}>0$).
The system now consists of independent two-site Hubbard models on disconnected rungs of the bilayer, each rung described by 
 \begin{align}
  H &=
   -   t_\perp \sum_{\sigma} 
      \left( \cdag_{A\sigma} \cccc_{B\sigma} + h.c. \right)
   \nonumber\\&{}~~~
  - \mu \sum_{i\sigma} n_{i\sigma}
  - |U| \sum_{i} (n_{i\up} - 1/2) (n_{i\dn} - 1/2)   
   \label{TwoSiteHubHamiltomian}
  \end{align}
where $i=A,B$ distinguishes the two sites.  Each two-site system can be occupied by $N=0,1,2,3,$ or $4$ fermions.  Exact diagonalization in the basis of 16 Fock states shows that the lowest-energy state in each subspace of fermion number $N$ is
  \begin{align}
   E_0 &= -\half \abs{U} + 2\mu			   	\nonumber\\
   E_1 &= -t_\perp + \mu		   						\nonumber\\
   E_2 &= -\sqrt{ \tfrac{1}{4} \abs{U}^2 + 4t_\perp^2}	\nonumber\\
   E_3 &= -t_\perp - \mu		   						\nonumber\\
   E_4 &= -\half \abs{U} - 2\mu			   	
      .
  \end{align}
Fig.~\ref{AtomicLimitEnergyLevels} shows the energy levels 
as a function of $U$ (for $\mu=0$).  The ground state is always at half-filling ($N=2$, black curve).  

For $\abs{U}/t_\perp < 2$, the cheapest excitation out of the $N=2$ subspace is a fermionic excitation, i.e., adding or removing one fermion to get to $N=1$ or $N=3$ (red line).  In this regime the model might be said to be a ``Fermi insulator,'' with a single-particle gap 
  \begin{align}
	E_g = \min\left( E_3 - E_2, E_1 - E_2 \right).
  \end{align}
For $\abs{U}/t_\perp > 2$, the cheapest excitation out of the $N=2$ subspace is a bosonic excitation, i.e., adding or removing a pair of fermions to get to the $N=0$ or $N=4$ subspace (blue line).  In this regime the model is a ``Bose insulator,'' with a two-particle gap
  \begin{align}
	\omega_\text{pair} = \min\left( E_4 - E_2, E_0 - E_2 \right).
  \end{align}

\section{Appendix C:  Atomic Limit plus Hopping}

We now consider the effects of a small in-plane hopping $t\ll t_\perp$ on the atomic limit results.

\newcommand\updn{\up\dn}
{\bf Corrections to ground state:}
For $t=0$, the half-filled ground state of the system can be schematically written 
as $\ket{\dots 2222 \dots}$.  Here each ``2'' represents the fact that there are 
two fermions on every rung in the bilayer lattice when the density per site is $n=1$.
In fact, for each rung, "2" is a specific linear combination of six states 
$\ket{\updn,0}, \ket{0,\updn}, \ket{\sigma,\sigma'}$ with two fermions on one rung, 
with definite amplitudes that depend on $|U|/t_\perp$.
In the presence of finite in-plane hopping $t>0$, the ground state develops an admixture of states such as $\ket{\dots 2312 \dots}$ and $\ket{\dots 2402 \dots}$.  We will ignore these corrections.

{\bf Corrections to excited state with one extra particle:}
If $t=0$, the lowest single-particle excited state is of the form $\ket{\dots 2322 \dots}$, and the single-particle gap is 
  \begin{align}
	E_g (0) &= E_3 - E_2 
		= \sqrt{ \tfrac{1}{4} \abs{U}^2 + 4t_\perp^2} - t_\perp - \mu
		.
  \end{align}
If $t>0$, the lowest single-particle excited state is mainly a superposition of states such as $\ket{\dots 2322 \dots}$ and 
$\ket{\dots 2232 \dots}$, connected by hopping $t$.  
%
Thus we might expect that the new single-particle gap is 
  \begin{align}
	E_g(t) \approx E_g(0) - \alpha t 
	\label{eqA10}
  \end{align}
where $\alpha$ is a constant of order unity.
  
In the strong-coupling limit $\abs{U} \gg t_\perp$, $E_g \approx \abs{U}/2$, so we expect the single-particle gap to collapse to zero at about $t\sim \abs{U}$.  However, this is not an insulator-metal transition (as we see in the next section).
Rather, the collapse of the ``nominal'' single-particle gap represents a crossover from a BEC superfluid to a BCS superfluid regime.  This behavior, $t_\text{BEC-BCS} \propto \abs{U}$, is shown as a dashed line in Fig.~1(a) of the text.  

In the weak-coupling limit $\abs{U} \ll t_\perp$, the above approximation suggests a transition at $t \sim t_\perp$.  This is a crude estimate;
we actually know that the transition actually occurs at 
$t \approx 2t_\perp/9$ as $|U|$ vanishes, as shown earlier.

{\bf Corrections to excited state with two extra particles:}
If $t=0$, the lowest two-particle excited state is of the form $\ket{\dots 2422 \dots}$, and the two-particle gap is 
  \begin{align}
	\omega_\text{pair} (0) &= E_4 - E_2 
		= \sqrt{ \tfrac{1}{4} \abs{U}^2 + 4t_\perp^2} - \half \abs{U} - 2\mu
		.
  \end{align}
If $t>0$, a local two-particle excitation can hop from one rung to an adjacent rung in a two-step process. 
 The effective boson hopping is of the order of $t_\text{boson} = 4t^2/\abs{U}$.  
%
Hence the two-particle gap is reduced to 
  \begin{align}
	\omega_\text{pair} (t) \approx 	\omega_\text{pair} (0) - \alpha t_\text{boson}
  \end{align}
where $\alpha$ is a constant of order unity (not the same as in Eq.~\eqref{eqA10}).
In the limit $\abs{U} \gg t_\perp$, we have 
$
	\omega_\text{pair} (0) \approx 4t_\perp {}^2  / \abs{U}
$.
This suggests that the two-particle gap $\omega_\text{pair}$ falls to zero when
$
	4t_\perp {}^2  / \abs{U} - \alpha \times 	4t^2  / \abs{U} = 0
$,
i.e., when $t \sim t_\perp$.  Then the system becomes overrun by two-particle excitations and makes a transition from BI to BEC.  This estimate is crude, 
as it neglects a large number of corrections to the initial and final state.  Nevertheless, we expect that the general idea is correct, i.e., that for $\abs{U} \gg t_\perp$ the BI-BEC phase boundary in the $(t,U)$ plane has a vertical asymptote at some finite value of $t/t_\perp$.  This behavior is illustrated in Fig.~1(a) of the text.

   \begin{figure}[!htb]
   \vspace{3mm}
	 \includegraphics[width=0.6\columnwidth]{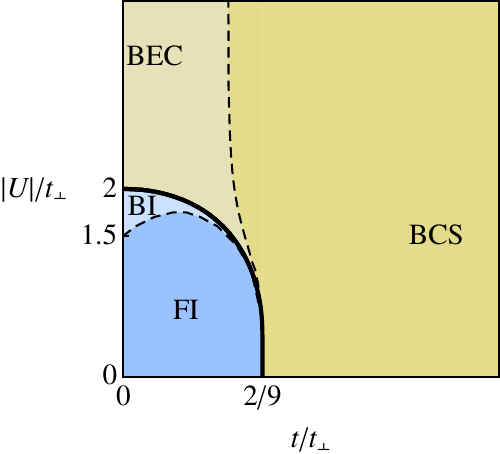} 
   \caption{
      \label{SchematicPhaseDiagramMFT-U-t}
			Schematic phase diagram from mean-field theory (MFT) at $T=0$. Note that the actual
			phase diagram in Fig.~\ref{PhaseDiagram}(a) is very different from MFT
			which is reliable only for small $|U|/t_\perp$; see text.
   }
   \end{figure}

\section{Appendix D: Pairing Susceptibility and Mean-Field Theory}
We next present a mean-field approach, which works best in the $|U|\ll t$ regime, and thus complements the atomic limit analysis presented in the previous two appendices.

{\bf Pairing instability:}
As discussed in the text, we first calculate the bare particle-particle channel susceptibility,
with center-of-mass momentum ${\bf q} = 0$, given by
	\begin{equation}
	\chi_0(\omega)\!=\! \frac{1}{N}\sum_{\kkk} \frac{1 - 2f_\kkk}{{2 \vare_\kkk - \omega - i0^+}}.
	\end{equation} 
Here $\vare_\kkk = \vare_\kkk^0 - \mu_H$, takes into account the Hartree shift $\mu_H = |U|(n-1)/2$
with the bare dispersion $\vare_\kkk^0$ given by Eq.~\eqref{tb-dispersion}.
$n$ is the density, $N$ the number of lattice sites and $f_\kkk$ the Fermi function.
The $\kkk$-sum is over the two bands $k_z =0,\pi$ and over the 2D Brillouin zone
$(k_x,k_y)$. 

In the metallic phase $\chi_0$ has the well-known $\ln \omega$ divergence of BCS theory.  
However, we concentrate here on the pairing instability in the insulating phase, where $\chi_0$ is finite.
The $T=0$ result  for the $n=1$ insulator, which corresponds to a choice of chemical potential $ -t_\perp + 3t < \mu < t_\perp - 6t $,
can be shown to be
  \begin{align}
	\chi_0 (\omega) &=
	\frac{1}{4t} \left[ 
			G_\text{tri} \left( \frac{\omega/2+\mu+t_\perp}{t}  \right) 
	-		G_\text{tri} \left( \frac{\omega/2+\mu-t_\perp}{t}  \right)
		\right] 
  \end{align}
We are able to write the susceptibility $\chi_0$ in terms of a single-particle
Green's function because we are looking only at pairs with total momentum ${\bf q} = 0$,
built up from two single-particle excitations with momenta $\pm \kkk$, and $\vare_\kkk^0 = \vare_{-\kkk}^0$.
We also find that similar expressions hold for the $n=0$ and $n=2$ insulating states.  

We then calculate the pairing susceptibility in the ladder approximation,
	\begin{align}
	\chi(\omega) &= \frac{1}{\chi_0^{-1}(\omega)  - |U|}.
	\end{align}
We find the critical in-plane hopping $t_c$ in MFT by locating the value of $t$ at which $\chi_0(\omega=0)=1/|U|$, 
so that $\chi(\omega)$ diverges.


Figure~\ref{PhaseDiagramMu} shows the MFT phase diagram in the $(t,\mu)$ plane, for a fixed coupling $\abs{U}/t_\perp=1.8$.
In the insulating phase there is a finite pairing susceptibility $\chi_0$, which can lead to a pairing instability in the presence of attraction $\abs{U}$.  As $\abs{U}$ increases, more and more of the insulating lobe is converted into a BEC superfluid.

Figure~\ref{SchematicPhaseDiagramMFT-U-t} shows the MFT phase diagram in the $(t,U)$ plane at half-filling.  As $\abs{U}$ increases, the system undergoes a transition from an insulator (INS) to a superfluid (BEC), due to the pairing instability.
At the end of this Appendix we will discuss the limitations of MFT and why the MFT results in
Fig.~\ref{SchematicPhaseDiagramMFT-U-t} look so different from the more correct results shown in Fig.~1(a).

{\bf Ordered state:}
In the superconducting state, there is a finite pairing amplitude
$\Delta =(|U|/N) \sum_i \langle c_{i\up} c_{i\dn} \rangle$.
We solve the mean-field theory (MFT) equations 
   \begin{align}
   {1}/{|U|} &= N^{-1}\sum_{\kkk}  \tanh({E_\kkk}/{2T})/{2E_\kkk}
   ,
   \label{MFTGapEquation}
   \\
   (n-1)/{2} &= -N^{-1} \sum_{\kkk} ({\vare_\kkk}/{2E_\kkk}) \tanh({E_\kkk}/{2T})
   \label{MFTNumberEquation}
      .
   \end{align}
to find $\Delta$ and $\mu$, where
the Hartree-shifted dispersion $\vare_\kkk = \vare_\kkk^0 - \mu_H$
determines the Bogoliubov quasiparticle spectrum
$E_\kkk = \sqrt{\vare_\kkk {}^2 + \Delta^2}$. Using this we can calculate
the energy gap $E_g = \min E_\kkk$ in the DOS $N(\omega)$
for single-particle excitations.
(The $\kkk$-integrals over the 2D Brillouin of the triangular lattice are performed by reducing them to 1D integrals using the exact density of states (DOS) in terms of
the triangular lattice Green function.)
The BEC-BCS boundary (dotted black curves in Figs.~\ref{PhaseDiagramMuEff} and \ref{PhaseDiagramMu}) is determined by the criterion that the dispersion minimum occurs at $\kkk=\0$ in the BEC regime and at $\kkk\neq \0$ in the BCS regime.

{\bf Choice of chemical potential:}
We are interested in studying the superconductor-insulator transition tuned by hopping $t$, rather than by fermion density $n$.  
Thus, in the superconducting phase we choose the chemical potential $\mu$ (according to Eq.~\eqref{MFTNumberEquation}) such that the average density corresponds to half-filling, $n=1$.  
In the insulating phase, we choose $\mu$ such that the two-particle excitation gap is particle-hole-symmetric.  In other words, the energy cost of adding a pair of fermions, $\omega_{e_2}$, is equal to the energy cost of removing a pair of fermions, $\omega_{h_2}$.
This choice of $\mu$ corresponds to the dashed blue curve in Figs.~\ref{PhaseDiagramMuEff} and \ref{PhaseDiagramMu}, which bisects the $n=1$ insulator lobe, and passes through the tip of the lobe.  The quantities in Fig.~2 of the text are plotted for this choice of $\mu$.

Note that the single-particle excitation gap, $E_g$, is \emph{not} particle-hole-symmetric along this trajectory.  The value of $E_g$ plotted in Fig.~2 of the text corresponds to the smaller of the particle gap and the hole gap, $\min(E_e,E_h)$.  

If one attempts to choose $\mu$ to make $E_e=E_h$, one finds that $\omega_{e_2} \neq \omega_{h_2}$.  This $\mu(t)$ trajectory exits through the side of the lobe rather than the tip of the lobe, representing a density-tuned transition rather than a hopping-tuned one.

{\bf Strengths and Limitations of MFT:}
MFT allows us to track the behavior of many quantities (see Fig.~2 in the text) that are intuitively meaningful, but not accessible in DQMC or other methods.
For fixed, small $|U|$ and varying $t$, MFT gives a good idea of the general behavior of various observables. 
Specifically, we can understand the insulator-to-SC transition at fixed, small $U$, where a change in the band-structure 
(reduction in band gap) causes $\chi_0$ to increase beyond $1/\abs{U}$ and precipitates a pairing instability.

However, MFT gives unreliable results as a function of $|U|$, particularly at large $|U|$.
We see this most clearly from a comparison of the MFT phase diagram Fig.~\ref{SchematicPhaseDiagramMFT-U-t} with the 
phase diagram of Fig.~\ref{PhaseDiagram}(a), which incorporates strong coupling and DQMC inputs, namely the existence 
of a SIT at non-zero value of $t/t_\perp$ at intermediate and large values of $|U|$.
Even at weak coupling, the MFT phase boundary $\abs{U}_c (t)$ for the SIT is a decreasing function of $t$, 
whereas atomic limit and DQMC results suggest that it is an increasing function as shown in 
Fig.~\ref{PhaseDiagram}(a). 

In MFT the interaction is decoupled exclusively in terms of the order parameter.  
Hence it fails to capture important correlations in the insulating state.
In the atomic limit ($t=0$), as $\abs{U}$ increases, the ground state wave-function adjusts itself to take advantage 
of the attraction to produce a binding energy.  There is no superfluid state at $t=0$ (as would have been predicted by MFT).
In fact the spurious MFT phase transition as a function of $|U|$ in Fig.~\ref{SchematicPhaseDiagramMFT-U-t} 
is actually akin to the Fermi-to-Bose crossover in the insulating state in Fig.~\ref{PhaseDiagram}(a).

\bibliographystyle{forprl}
\bibliography{sit}

\end{document}